# Electrically driven optical interferometry with spins in silicon carbide

## Authors

Kevin C. Miao[1], Alexandre Bourassa[1], Christopher P. Anderson[1,2], Samuel J. Whiteley[1,2], Alexander L. Crook[1,2], Sam L. Bayliss[1], Gary Wolfowicz[1], Gergő Thiering[3], Péter Udvarhelyi[4,3], Viktor Ivády[3,5], Hiroshi Abe[6], Takeshi Ohshima[6], Ádám Gali[3,7] & David D. Awschalom[1,2,8,*]

## Affiliations

[1] *Institute for Molecular Engineering, University of Chicago, Chicago, IL 60637, USA*
[2] *Department of Physics, University of Chicago, Chicago, IL 60637, USA*
[3] *Wigner Research Centre for Physics, Hungarian Academy of Sciences, PO Box 49, H-1525, Budapest, Hungary*
[4] *Department of Biological Physics, Loránd Eötvös University, Pázmány Péter sétány 1/A, H-1117 Budapest, Hungary*
[5] *Department of Physics, Chemistry and Biology, Linköping University, SE-581 83 Linköping, Sweden*
[6] *National Institutes for Quantum and Radiological Science and Technology, 1233 Watanuki, Takasaki, Gunma 370-1292, Japan*
[7] *Department of Atomic Physics, Budapest University of Technology and Economics, Budafoki út 8., H-1111, Budapest, Hungary*
[8] *Institute for Molecular Engineering and Materials Science Division, Argonne National Laboratory, Lemont, IL 60439, USA*

[*] email: awsch@uchicago.edu

## Abstract

Interfacing solid-state defect electron spins to other quantum systems is an ongoing challenge. The ground-state spin's weak coupling to its environment bestows excellent coherence properties, but also limits desired drive fields. The excited-state orbitals of these electrons, however, can exhibit stronger coupling to phononic and electric fields. Here, we demonstrate electrically driven coherent quantum interference in the optical transition of single, basally oriented divacancies in commercially available 4H silicon carbide. By applying microwave frequency electric fields, we coherently drive the divacancy's excited-state orbitals and induce Landau-Zener-Stückelberg interference fringes in the resonant optical absorption spectrum. Additionally, we find remarkably coherent optical and spin subsystems enabled by the basal divacancy's symmetry. These properties establish divacancies as strong candidates for quantum communication and hybrid system applications, where simultaneous control over optical and spin degrees of freedom is paramount.

## MAIN TEXT

## Introduction

Solid-state spins have risen to prominence as qubits with atom-like properties combined with the fabrication advantages offered by the solid-state host. In particular, the neutral divacancy (VV) defect in silicon carbide (SiC) has attracted considerable attention for its near-infrared optical addressability (*1*), highly coherent spin (*2*), and adaptability stemming from SiC polytypism (*3*). Multiple methods of quantum control (*1*,*4*-*6*) have been demonstrated using the VV electron spin, while the SiC host material can be doped (*7*) and fabricated into nanostructures (*8*,*9*). Furthermore, *c*-axis oriented VVs possess a high-fidelity spin-to-photon interface (*10*), and exhibit large dc Stark shifts under applied electric fields (*7*,*11*) and strain (*12*).

While the ground state of solid-state electron spins can be coherently controlled with acoustic (*5*,*13*,*14*) and electric (*4*) fields, the excited-state orbitals of these electrons can exhibit significantly larger acoustic (*12*,*15*,*16*) and electric (*7*,*11*,*17*,*18*) couplings, leading to enhanced interactions. In this work, we employ the ground- and excited-state electronic orbital levels of single, basally oriented *kh* VVs in the 4H polytype of SiC to implement an optical two-level system (TLS). We demonstrate that this TLS, which has near-lifetime-limited coherence, can couple to electric fields, inducing a controllable $\sigma_z$ (Pauli-Z) interaction in its Hamiltonian. This enables us to use ac electric fields to modulate the intensity and frequency profile of the *kh* VV resonant absorption spectrum between the levels of the TLS, inducing Landau-Zener-Stückelberg (LZS) interference fringes. Most notably, the electric field coupling strength to the excited-state orbital is large enough that we observe classically driven coherent interactions with up to 15 microwave photons even in the absence of a microwave resonator. We show the coupling can be driven at a wide range of frequencies by demonstrating the TLS's complex spectral response under two simultaneous electric field drives at distinct frequencies. Finally, we show that alongside the coherent optical interface, the *kh* VV's $C_{1h}$ symmetry leads to a ground-state spin with long dephasing times and magnetically drivable spin rotations that span the eigenbasis. These results demonstrate extensive optical and spin control of *kh* VVs, and pave the way for advanced quantum optics applications demanding high coherence across both photonic and spin subsystems.

## Results

We first outline the *kh* VV's physical configuration and associated orbital and spin energy levels. The *kh* VV defect in 4H-SiC consists of a carbon vacancy $V_C$ adjacent to a silicon vacancy $V_{Si}$ in the basally oriented *kh* configuration, where *k* and *h* refer to quasi-cubic and hexagonal local environments of the $V_C$ and $V_{Si}$ sites, respectively. Dangling bonds from the six neighboring atoms form a localized $C_{1h}$ symmetry system, with a spin-1 orbital singlet ground state and orbital doublet excited state in the SiC electronic bandgap (Fig. 1A,B). The resulting defect quantization axis is tilted at an angle of 71° with respect to the crystal *c*-axis. At 5 K, we take scanning confocal images of a 4H-SiC sample using 905 nm excitation (see Methods), revealing isolated emitters in the 10 µm wide gap of an on-chip capacitor (Fig. 1C). The emission spectrum of these isolated emitters shows about 10% of the emission in a zero-phonon line (ZPL) near 1079 nm, along with a redshifted phonon sideband (PSB) extending out to 1200 nm (Fig. 1A, inset), confirming the optical signature as *kh* VVs (*1*,*3*).

The $C_{1h}$ symmetry manifests in the *kh* VV ground-state spin sublevels as longitudinal and transverse zero-field splittings $D_{GS}$ and $E_{GS}$, producing spin eigenstates $\left\{|+\rangle = \frac{|+1\rangle + |-1\rangle}{\sqrt{2}}, |0\rangle, |-\rangle = \frac{|+1\rangle - |-1\rangle}{\sqrt{2}}\right\}$ in the $S_z = \{|+1\rangle, |0\rangle, |-1\rangle\}$ basis at low magnetic field (see Methods). We label this reduced symmetry spin triplet as ${}^3\tilde{A}_2$. Zero First-Order Zeeman (ZEFOZ) transitions emerge (see Methods), suppressing magnetic field noise. Additionally, magnetically driven spin transitions arise between $|0\rangle \leftrightarrow |\pm\rangle$ as well as $|+\rangle \leftrightarrow |-\rangle$, which we describe in detail later. In the excited state, the $C_{1h}$ symmetry presents itself as a transverse crystal field, producing two widely separated ${}^3A'$ and ${}^3A''$ orbitals (Fig. 1B). We do not observe emission from ${}^3A''$ at 5 K, suggesting the presence of internal conversion causing fast relaxation from ${}^3A''$ to ${}^3A'$. Each orbital has three spin sublevels, and the fine structure of each orbital is also subject to longitudinal and transverse zero-field splittings $D_{ES}$ and $E_{ES}$, which produce spin mixing similar to that in the ground state. Thus, we label the spin sublevels of ${}^3A'$ as $|A'_0\rangle$, $|A'_+\rangle$, $|A'_-\rangle$, analogous to $|0\rangle$, $|+\rangle$, $|-\rangle$ of ${}^3\tilde{A}_2$, respectively.

In order to map the fine structure of ${}^3A'$, we perform spin-dependent photoluminescence excitation (PLE) spectroscopy (see Methods) on a single *kh* VV in a charge-depleted environment, in which photoinduced charge dynamics of impurity sites are suppressed and spectral diffusion is reduced (*7*). We identify a central resonance corresponding to the $|0\rangle \leftrightarrow |A'_0\rangle$ transition, as well as two weaker detuned resonances primarily corresponding to $|+\rangle \leftrightarrow |A'_+\rangle$ and $|-\rangle \leftrightarrow |A'_-\rangle$ (Fig. 2A). From these resonance locations, we estimate the excited state longitudinal and transverse excited-state zero-field splittings, $D_{ES}$ and $E_{ES}$, to be +970 MHz and −483 MHz, respectively. The strikingly large magnitude of $E_{ES}$ suggests a significant change in electronic wavefunction distribution between the ground and excited state (see Supplementary Information). Furthermore, the spin selective nature of these transitions can be used for high-fidelity readout of the ground-state spin. While residual spectral diffusion of the optical transition frequency is observed (see Supplementary Information), we can sum over individual scans with the wandering compensated, allowing us to extract a 21.0(2) MHz full width at half maximum of the Lorentzian lineshape (Fig. 2A, inset).

We probe the excited-state dynamics of single *kh* VVs by performing time-correlated fluorescence measurements (see Methods). Optical excitation of $|0\rangle \leftrightarrow |A'_0\rangle$ and $|+\rangle \leftrightarrow |A'_+\rangle$ reveals coherent control of both transitions (Fig. 2B). A fit of the data to an integrated Bloch equation model reveals longitudinal and transverse relaxation times, as well as the spin relaxation rate under illumination ($T_1$, $T_2$, $\Gamma$, respectively) (see Methods). On the timescale of the experiment, minimal spin relaxation under illumination for $|0\rangle \leftrightarrow |A'_0\rangle$ indicates that it may be cycling, while $|+\rangle \leftrightarrow |A'_+\rangle$ is non-cycling and can be used for efficient spin pumping. For both transitions, the optical coherence $T_2$ approaches the lifetime limit of $2T_1$. We ascribe this to an exponential suppression of orbital mixing given the large crystal field splitting, since the energy difference between the orbitals is expected to be larger than $k_BT$. The near-lifetime-limited coherence of both optical transitions suggests that the spectral linewidth obtained through PLE spectroscopy may still be broadened by optical driving (*19*). Furthermore, the fidelity of coherent phenomena, such as interferometry and interaction gates in hybrid systems, benefits from long coherence times. In particular, the profile of the LZS interference spectrum depends on optical coherence time, which we show in detail later.

With the optical properties characterized, we can isolate a TLS in the energy structure and demonstrate LZS interference under an ac electric field drive. For our TLS, we use $|0\rangle \leftrightarrow |A_0'\rangle$ for its brightness and all-optical preparation sequence. LZS interference arises when the TLS is repeatedly brought through an avoided crossing diabatically, while a Stückelberg phase is acquired between each crossing (*20*). Optical driving of the TLS produces an avoided crossing, while the Stark effect allows for controlled passage through that crossing. We couple highly confined electric fields between the on-chip planar capacitor plates (Fig. 1C) to single *kh* VVs, yielding dc Stark shifts of the VV's excited-state orbital levels that are resolvable through PLE spectroscopy (see Supplementary Information). We then apply a GHz-frequency ac electric field drive with zero dc bias concurrently with the resonant excitation. We monitor the PLE spectrum as we vary the amplitude $|\boldsymbol{F}|$ of a sinusoidal electric field drive with frequency $\omega = 2\pi \times 700$ MHz. At zero amplitude, only the bare resonance of $|0\rangle \leftrightarrow |A_0'\rangle$ is observed at zero laser detuning. With increasing amplitude, we produce multiphoton resonances at laser detunings equal to integer multiples of the drive frequency, culminating in classical interactions with up to 15 microwave photons when the laser is detuned by $\delta = 15 \times 2\pi \times 700$ MHz (Fig. 3A).

With the proper rotating frame transformations (see Methods), our observations can be explained by the Hamiltonian of the TLS in the rotating frame:

$$\frac{H''(t)}{\hbar} = \sum_{n=-\infty}^{\infty} \frac{\Delta_n}{2}\left(e^{-in\omega t}\sigma_+ + e^{in\omega t}\sigma_-\right) + \frac{\delta}{2}\sigma_z \tag{1}$$

$$\Delta_n = \Omega J_n\left(\frac{\mathcal{A}}{\omega}\right)$$

$$\sigma_\pm = \frac{1}{2}\left(\sigma_x \pm i\sigma_y\right)$$

where $\delta$ is the optical detuning, $\Omega$ is the optical Rabi frequency, $J_n(x)$ are Bessel functions of the first kind, $\mathcal{A}$ is the amplitude of the induced Stark shift from the ac electric field, and $\omega$ is the ac electric field frequency. This Hamiltonian leads to LZS interference and multiphoton interactions where the bare optical resonance absorbs or emits quanta of energy resonant with the drive. Integrating the photoluminescence within the baseband resonance shows an approximate $\left|J_n\left(\frac{\mathcal{A}}{\omega}\right)\right|$ dependence, which is expected in the strong optical driving regime, $T_1T_2\Omega^2 \gg 4\pi^2$ (see Supplementary Information). Together, these results demonstrate coherent control of a single *kh* VV's orbital levels with microwave-frequency electric fields, which can enable a high-fidelity microwave-to-optical interface when driven with single quanta of excitation.

We then turn our attention to bichromatic ac electric field driving enabled by the non-resonant capacitive structure. We apply drives $H_1 = \mathcal{A}_1\cos(\omega_1 t)\sigma_z$ and $H_2 = \mathcal{A}_2\cos(\omega_2 t + \varphi)\sigma_z$, with amplitudes of both drives tuned to the first Bessel zero of the baseband resonance, $\mathcal{A}_1/\omega_1 = \mathcal{A}_2/\omega_2 \approx 2.4048$. We select two arbitrary, commensurable drive frequencies, $\omega_1 = 2\pi \times 1$ GHz and $\omega_2 = 2\pi \times 2$ GHz, both of which couple to the *kh* VV with similar strengths given the broadband nature of the capacitor (see Supplementary Information). The steady state of the two-level system can then be described by higher-dimensional generalized Bessel functions (*21*).

Specifically, the relative phase of the two drives can be tuned to break time reversal symmetry (*22*) and produce phase-dependent coherent destruction of tunneling. PLE spectroscopy taken while varying the relative phase $\varphi$ reveals a phase-dependent structure well-matched to simulation (Fig. 3B), indicating coherent coupling to electric fields at two frequencies spanning an octave. These results suggest that topological systems (*23*) can be simulated using polychromatic LZS phenomena in an electrically driven *kh* VV.

Finally, we demonstrate the existence of a particularly versatile and coherent ground-state spin system in single *kh* VVs. At zero effective magnetic field (see Methods), $\mathbf{B}_{\mathrm{eff}} = \vec{0}$, the mixed spin basis and resulting vanishing first derivative of the energy dispersion relation enable magnetically driven transitions between all three spin states as well as long spin dephasing times (Fig. 4A). We first show the electron spin can span its eigenbasis by magnetically driving the $|0\rangle \leftrightarrow |+\rangle$ and $|+\rangle \leftrightarrow |-\rangle$ transitions at $D_{\mathrm{GS}} + E_{\mathrm{GS}}$ = 1.352373 GHz and $2E_{\mathrm{GS}}$ = 36.839 MHz, respectively (see Supplementary Information). Using resonant spin-selective readout, we observe Rabi oscillations marked by high photoluminescence contrast (Fig. 4B,C). We then perform Ramsey interferometry on single *kh* VVs prepared into $|\psi_0\rangle = \frac{1}{\sqrt{2}}\left(|0\rangle + |+\rangle\right)$ at the ZEFOZ point, revealing a ground-state spin dephasing time of $T_2^*$ = 74 μs (Fig. 4D). Compared to *c*-axis VVs, which have ground-state spin dephasing times around 2 μs (*24*), the lengthened spin dephasing time can be attributed to the suppression of first-order magnetic field sensitivity of the ZEFOZ point. Residual second-order magnetic field sensitivity and first-order electric field, however, may ultimately limit $T_2^*$ (*25*,*26*). We use a Hahn-echo sequence to suppress higher-order magnetic field noise, allowing us to observe a spin coherence time of $T_2$ = 222 μs at the ZEFOZ point (Fig. 4E), comparable to low-field coherence times reported in *c*-axis VVs (*2*). A Gaussian decay envelope suggests that bath $^{29}$Si and $^{13}$C nuclear spins are a main source of spin decoherence, and that these contributions can be reduced by higher-order dynamical decoupling sequences such as XY-*n* (*27*).

**Discussion**

Novel interactions between optical and microwave photons in a solid-state system can be realized by leveraging the coherent coupling between electric fields and the excited-state orbitals of the *kh* VV. A bevy of microwave resonator engineering techniques allows for adaptations to various applications utilizing electric field coupling. For example, recent advances in high kinetic inductance superconducting resonators (*28*) can be leveraged to significantly enhance zero-point fluctuations of the electric field. Consequently, protocols inspired by both atomic physics and circuit quantum electrodynamics can be explored, including qubit-assisted resonator cooling where the optical transitions are driven in the resolved-sideband regime (*29*), and cavity-enhanced quantum state readout (*30*). Furthermore, the relative ease of confining electric fields allows for independent interrogation of isolated, yet spatially proximal VVs, enabling a scalable architecture for multi-qubit interactions (*31*). Meanwhile, remote entanglement applications requiring high Hong-Ou-Mandel interference visibility will benefit from the lifetime-limited coherence of the optical transitions (*32*). Collectively, these properties make *kh* VVs a versatile and promising platform to explore spin and optical dynamics coupled to an assortment of quantum control mechanisms.

## Methods

### Sample preparation and experimental setup

We create samples from a 4” commercially available wafer where 20 μm of high-purity i-type SiC ($[V_C]$,$[V_{Si}]$,[N] < $1 \times 10^{15}$ $cm^{-3}$) is epitaxially grown on a 4° off-axis miscut of the Si face of a high-purity semi-insulating SiC substrate (Norstel AB, serial number A3177-14). We dice 5 mm by 5 mm chips and perform electron irradiation with 2 MeV electrons at a dose of $3 \times 10^{12}$ $e^{-}/cm^2$ to create both $V_C$ and $V_{Si}$. Subsequent annealing at 850 °C for 30 minutes in Ar forms VV complexes uniformly throughout the entire sample. Planar capacitors with a 10 μm gap width and wires of width 10 μm made of Ti/Au are then patterned on the sample surface using electron beam lithography. Samples are cooled to 5 K in a closed-cycle cryostat (Montana Instruments, Cryostation s100).

Using a confocal microscopy setup, we perform confocal raster scans using off-resonant excitation at 905 nm (QPhotonics, QFLD-905-200S) or 975 nm (Thorlabs, BL975-SAG300) focused through a microscope objective (Olympus, LCPLN100XIR). We detect emitted photons with >80% quantum efficiency using a low jitter, low dark count superconducting nanowire single photon detector (SNSPD) (Quantum Opus, Opus One). Electrical pulses from the SNSPD are converted to TTL and counted using a data acquisition module (National Instruments, PCI-6259). We apply a bandpass filter with a passband of 1064-1150 nm in order to selectively detect emission from *kh* divacancies while nearly completely rejecting emission from other defects and divacancy orientations. We perform photoluminescence excitation (PLE) by scanning a narrow-line tunable laser (Toptica Photonics, DL pro) and collecting from the phonon sideband of the *kh* divacancies using a tunable longpass filter (Semrock, TLP01-1116-25x36). We initialize the spin with a pulse of off-resonant light, followed by spin manipulation using MW and RF pulses. A pulse of resonant light performs spin-dependent readout. Spin driving with on-chip planar wires is achieved using a signal generator (SRS, SG396) modulated by an arbitrary waveform generator (Tektronix, AWG5014C, for $|0\rangle \leftrightarrow |\pm\rangle$ transitions) or an IQ modulator (Analog Devices, LTC5598, for $|+\rangle \leftrightarrow |-\rangle$ transition). LZS interferometry spectra are produced with microwave-frequency electric fields applied during the resonant readout pulse. We apply microwaves generated in a similar fashion for spin driving across the on-chip capacitor, producing electric field magnitudes as high as 2 MV/m across the capacitor gap. Time-resolved photon arrival measurements are performed by time-tagging the SNSPD output pulses (Swabian Instruments, Time Tagger 20). We initialize the spin to $|0\rangle$ and perform MW rotations to prepare $|+\rangle$, followed by a laser pulse resonant with $|0\rangle \leftrightarrow |A_0'\rangle$ or $|+\rangle \leftrightarrow |A_+'\rangle$ for 80 ns. The arrival times of emitted photons in the PSB are time-tagged relative to the start of the resonant laser pulse. Approximately 5 Mcts are collected over 60 minutes and histogrammed into 1 ns bins to observe time-resolved dynamics.

### Spin Hamiltonian

The Hamiltonian of the divacancy ground state spin-1 system is

$$H/h = D\left(\hat{S}_z^2 - \frac{S(S+1)}{3}\right) + E\left(\hat{S}_+^2 + \hat{S}_-^2\right) + g\mu_B \mathbf{B}\cdot\hat{\mathbf{S}} + \sum_i \hat{\mathbf{S}}\cdot\mathbf{A}_i\cdot\hat{\mathbf{I}}_i \qquad (2)$$

where $D$ and $E$ are the ground-state longitudinal and transverse zero-field splittings, respectively, $\mu_B$ is the Bohr magneton, $g$ is the electron spin g-factor, $\mathbf{B}$ is the external magnetic field, $\mathbf{A}_i$ is the hyperfine tensor of nuclear spin $i$, and $\hat{\mathbf{I}}_i$ is the nuclear spin operator vector of nuclear spin $i$. Zeeman interactions of the nuclear spin are neglected.

For simplicity, we will consider a system with one nuclear spin. Assuming $B_x$ and $B_y$ are negligible, diagonalization of this Hamiltonian yields the following eigenenergies:

$$\mathrm{eigval}\left(H/h\right)=\begin{cases} D+\sqrt{C_+^2+E^2}, & |1\rangle \\ D+\sqrt{C_-^2+E^2}, & |2\rangle \\ D-\sqrt{C_+^2+E^2}, & |3\rangle \\ D-\sqrt{C_-^2+E^2}, & |4\rangle \\ 0, & |5\rangle \\ 0, & |6\rangle \end{cases} \tag{3}$$

and corresponding (unnormalized) eigenstates with a nuclear spin basis $I_z=\{|\uparrow\rangle,|\downarrow\rangle\}$:

$$\mathrm{eigvec}\left(H/h\right)=\begin{cases} \frac{E}{C_+ +\sqrt{C_+^2+E^2}}|+1\uparrow\rangle+|-1\uparrow\rangle, & |1\rangle \\ \frac{E}{C_- +\sqrt{C_-^2+E^2}}|+1\downarrow\rangle+|-1\downarrow\rangle, & |2\rangle \\ \frac{E}{C_+ -\sqrt{C_+^2+E^2}}|+1\uparrow\rangle+|-1\uparrow\rangle, & |3\rangle \\ \frac{E}{C_- -\sqrt{C_-^2+E^2}}|+1\downarrow\rangle+|-1\downarrow\rangle, & |4\rangle \\ |0\uparrow\rangle, & |5\rangle \\ |0\downarrow\rangle, & |6\rangle \end{cases} \tag{4}$$

where $C_\pm = B_z \pm \frac{A_{zz}}{g\mu_B}$. When $C_\pm = 0$, the effect of $B_z$ is to negate the hyperfine interaction for a particular nuclear spin orientation. For example, if we consider the system when the nuclear spin is $|\uparrow\rangle$ and $C_+ = 0$, the electron spin Hamiltonian becomes equivalent to one with no nuclear spin interactions or external magnetic field, which allows us to label the effective external magnetic field as $B_{z,\mathrm{eff}} = B_z + \frac{A_{zz}}{g\mu_B} = 0$. At $C_+ = 0$ ($S_y'$), the first derivative of the energy difference $\nu$ between $|1\rangle$ ($|2\rangle$) and $|4\rangle$ ($|3\rangle$) as a function of $B_z$ is zero, $\frac{d\nu}{dB_{z,\mathrm{eff}}} = 0$. Similar expressions can be derived for $B_x$ and $B_y$, showing that $\frac{d\nu}{dB_{x,\mathrm{eff}}} = \frac{d\nu}{dB_{y,\mathrm{eff}}} = 0$, and that the field is $\mathbf{B}_{\mathrm{eff}} = \mathbf{0}$ in order to fulfill all these conditions. The vanishing first derivative of the energy dispersion, by definition, produces a ZEFOZ transition.

By operating at the ZEFOZ point, we select out one of the two nuclear spin states. The effective Hamiltonian can be expressed in the $S_z = \{|+1\rangle, |0\rangle, |-1\rangle\}$ basis,

$$H_{\text{ZEFOZ}} / h = \begin{bmatrix} D & 0 & E \\ 0 & 0 & 0 \\ E & 0 & D \end{bmatrix} \tag{5}$$

which can be diagonalized

$$H'_{\text{ZEFOZ}} / h = U^{\dagger} H_{\text{ZEFOZ}} U / h = \begin{bmatrix} D+E & 0 & 0 \\ 0 & 0 & 0 \\ 0 & 0 & D-E \end{bmatrix} \tag{6}$$

$$U = \frac{1}{\sqrt{2}} \begin{bmatrix} 1 & 0 & 1 \\ 0 & \sqrt{2} & 0 \\ 1 & 0 & 1 \end{bmatrix}$$

where the eigenenergies of the three electron spin states are 0 and $D \pm E$, with corresponding eigenstates $|0\rangle$ and $\frac{1}{\sqrt{2}}\left(|+1\rangle \pm |-1\rangle\right)$. Spin-1 matrices in this basis can be expressed as

$$S'_x = \begin{bmatrix} 0 & 1 & 0 \\ 1 & 0 & 0 \\ 0 & 0 & 0 \end{bmatrix} \quad S'_y = \begin{bmatrix} 0 & 0 & 0 \\ 0 & 0 & 1 \\ 0 & 1 & 0 \end{bmatrix} \quad S'_z = \begin{bmatrix} 0 & 0 & 1 \\ 0 & 0 & 0 \\ 1 & 0 & 0 \end{bmatrix} \tag{7}$$

Thus, spin rotations can be achieved using the non-zero matrix elements (and their Hermitian conjugates) $\langle 0|S'_x|+\rangle$, $\langle 0|S'_y|-\rangle$, and $\langle -|S'_z|+\rangle$. Experimentally, we can realize $S'_x$ and $S'_z$ rotations, while $S'_y$ rotations induce significant off-resonant $S'_x$ rotation given the defect orientation with respect to the planar drive wire.

**LZS Hamiltonian**

The Hamiltonian of the unperturbed optical transition can be considered as a two-level system (TLS) with a ground and excited state $|g\rangle = |0\rangle$ and $|e\rangle = |A'_0\rangle$, respectively. In the $\sigma_z = |e\rangle\langle e| - |g\rangle\langle g|$ basis, this Hamiltonian can be expressed as:

$$H(t) / \hbar = \frac{\Omega \cos\left(\omega_{\text{opt}} t\right)}{2} \sigma_x + \frac{\omega_0}{2} \sigma_z \tag{8}$$

where $\Omega$ is the optical Rabi frequency, $\omega_{\text{opt}}$ is the laser frequency, and $\omega_0$ is the optical resonance frequency. In the rotating frame of the TLS, a time-independent Hamiltonian arises:

$$H / \hbar = \frac{\Omega}{2} \sigma_x + \frac{\delta}{2} \sigma_z \tag{9}$$

where $\delta = \omega_{\text{opt}} - \omega_0$ is the laser detuning from the resonance frequency. This system is analogous to the unperturbed TLS presented in Shevchenko, *et al.* (*33*), with the substitutions $\Delta \to -\Omega$ and $\varepsilon \to -\delta$.

The addition of a longitudinally coupled ac electric field drive of amplitude $\mathcal{A}$ and frequency $\omega$ corresponds to an additional time-dependent Hamiltonian,

$$H_{\text{a.c.}}(t)/\hbar = \frac{\mathcal{A}\cos(\omega t)}{2}\sigma_z \tag{10}$$

so that

$$H'(t)/\hbar = \frac{H + H_{\text{a.c.}}(t)}{\hbar} = \frac{\Omega}{2}\sigma_x + \frac{\delta + \mathcal{A}\cos(\omega t)}{2}\sigma_z \tag{11}$$

Applying a transformation (33) into a frame rotating with $H_{\text{a.c.}}(t)$ to $H'(t)$ yields

$$H''(t)/\hbar = \frac{\delta}{2}\sigma_z + \sum_{n=-\infty}^{\infty}\frac{\Delta_n}{2}\left(e^{-in\omega t}\sigma_+ + e^{in\omega t}\sigma_-\right) \tag{12}$$

where $\Delta_n = \Omega J_n\left(\frac{\mathcal{A}}{\omega}\right)$ is the effective optical Rabi frequency under ac electrical drive.

## Optical Bloch equation model

We model the *kh* divacancy optical structure as a three-level system in the presence of spin relaxation under illumination. In addition to the ground and excited states $|g\rangle$ and $|e\rangle$, we also consider a trapping state $|s\rangle$ representing non-interacting ground-state spin states. Furthermore, we introduce three relaxation channels in the form of Lindblad operators $C_n$: radiative decay, $C_1 = \frac{1}{\sqrt{T_1}}|g\rangle\langle e|$; pure dephasing, $C_2 = \frac{1}{\sqrt{T_2^*}}\left(|e\rangle\langle e| - |g\rangle\langle g|\right)$; and spin relaxation under illumination, $C_3 = \sqrt{\Gamma}|s\rangle\langle e|$. Our Hamiltonian takes the form

$$H(t)/\hbar = \frac{\Omega(t)}{2}\left(|g\rangle\langle e| + |e\rangle\langle g|\right) + \delta|e\rangle\langle e| \tag{13}$$

We then use `mesolve` provided by the QuTiP Python package, which solves the Lindblad master equation:

$$\dot{\rho}(t) = -\frac{i}{\hbar}\left[H(t), \rho(t)\right] + \frac{1}{2}\sum_n\left(2C_n\rho C_n^\dagger - \left\{C_n^\dagger C_n, \rho\right\}\right) \tag{14}$$

where $\rho$ is the density matrix, $[A, B] = AB - BA$ and $\{A, B\} = AB + BA$. Fitting to the data allows us to extract $T_1$, $T_2 = \left((2T_1)^{-1} + T_2^{*-1}\right)^{-1}$ and $\Gamma^{-1}$.

## DFT methodology

We use a plane wave basis set and PAW atomic potentials as implemented in VASP. The 4H-SiC supercell includes 576 atoms and embeds a single *hh* or *kh* divacancy. We use two exchange correlation functionals in our calculations, either the semi-local PBE functional or the HSE06 hybrid functional. The calculations are performed with a 420 eV plane wave cutoff energy using Γ-point sampling of the Brillouin zone. The excited-state electronic structure and geometry are calculated by constraint occupation of states. The splitting between the excited state branches is determined from the splitting of the Kohn-Sham energies of the $a'$ and $a''$ single particle states in a $a'_{a1}(1)a'_{e}(1.5)a''_{e}(1.5)$ single particle configuration. The relaxation path between the $^3A''$ and the $^3A'$ states is determined by nudged elastic band calculations. To calculate the permanent polarization of states, we use the VASP implementation of the Berry phase theory of polarization.

**Acknowledgments**

**General**: We thank M. Fukami, B. Zhou, F. J. Heremans, V. Dobrovitski, and A. Clerk for helpful discussions, and B. Diler for machining assistance.

**Funding:** K.C.M., A.B., C.P.A., S.J.W., A.L.C., S.L.B., G.W., and D.D.A. were supported by AFOSR FA9550-14-1-0231 and FA9550-15-1-0029, DARPA D18AC00015KK1932, NSF EFRI EFMA-1641099, and ONR N00014-17-1-3026. G.T., P.U., V.I., and A.G. were supported by National Research, Development and Innovation Office in Hungary (NKFIH) Grant Nos. 2017-1.2.1-NKP2017-00001 (National Quantum Technology Program) and NVKP 16-1-2016-0043 (NVKP Program) as well as Grant Nos. NN127902 (EU QuantERA Nanospin consortial project) and KKP129866 (Quantum-coherent materials project within National Excellence program), and from the EU Commission (ASTERIQS project with Grant No. 820394). V.I. was supported by the MTA Premium Postdoctoral Research Program and the Knut and Alice Wallenberg Foundation through WBSQD2 project (Grant No. 2018.0071). H.A. and T.O. were supported by JSPS KAKENHI 17H01056 and 18H03770. This work made use of the Pritzker Nanofabrication Facility of the Institute for Molecular Engineering at the University of Chicago, which receives support from Soft and Hybrid Nanotechnology Experimental (SHyNE) Resource (NSF ECCS1542205), a node of the National Science Foundation's National Nanotechnology Coordinated Infrastructure.

**Author contributions:** K.C.M. conceived and demonstrated the optical interference properties with the help of S.J.W., performed VV experiments, and wrote the manuscript. K.C.M., A.B., and A.L.C. developed the scanning confocal microscopy setup. C.P.A. annealed the SiC sample and fabricated planar capacitors and wires on the sample. K.C.M. and S.L.B. performed coherent optical control experiments. G.W. performed CCE modeling of *kh* VVs. G.T., P.U., V.I., and Á.G. carried out DFT calculations of the optical structure and response to electric field perturbation. H.A. and T.O. performed electron irradiation of the SiC samples. D.D.A. advised on all efforts. All authors contributed to the discussions and preparation of the manuscript.

**Competing interests:** The authors declare that they have no competing financial interests.

**Data and materials availability:** The datasets generated and analyzed during the current study are available from the corresponding author upon reasonable request.

## Figures

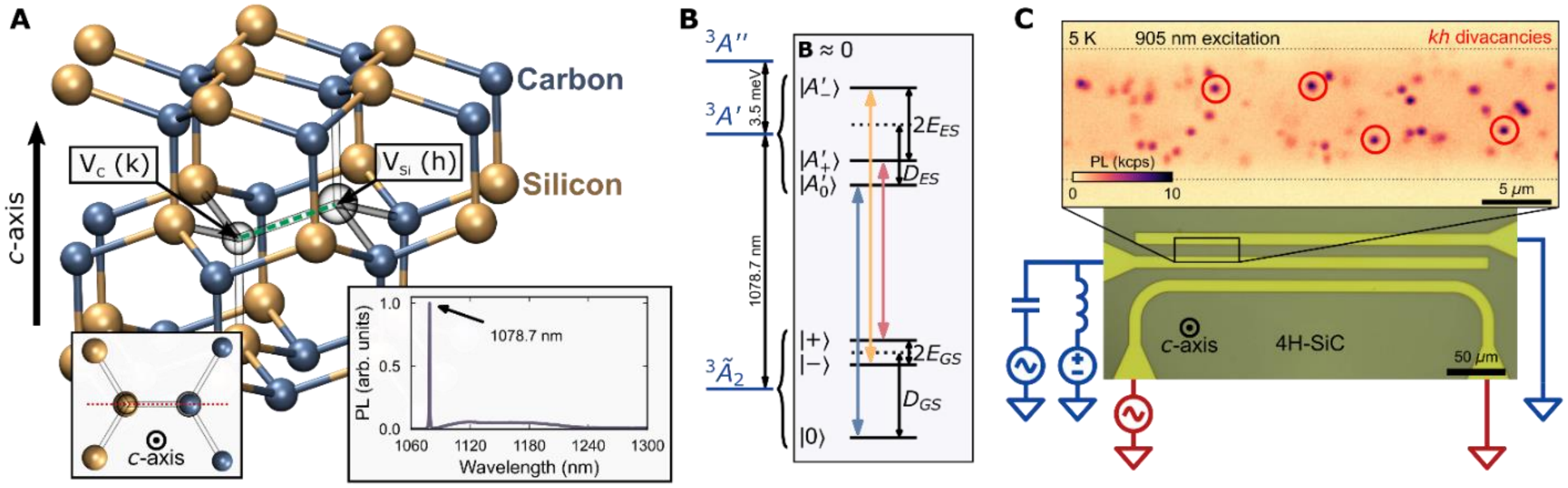

**Fig. 1. Single *kh* VVs in commercially available 4H-SiC.** (**A**) Lattice configuration of *kh* VVs in 4H-SiC. The defect axis is indicated by the green dashed line. Inset, left: Local atomic configuration around the *kh* VV showing $C_{1h}$ symmetry. Inset, right: Emission spectrum of a single *kh* VV. (**B**) Energy diagram of the *kh* VV. The spin sublevels mix due to the effect of transverse zero-field splitting $E_{GS/ES}$, causing $|\pm1\rangle$ to become $|\pm\rangle$ near zero external magnetic field. Spin-selective optical transitions (blue, yellow, and red arrows) enable spin state readout. (**C**) Optical image of the 4H-SiC sample with a lithographically patterned capacitor and wire. Inset: Scanning confocal image of the marked region between the coplanar capacitor plates using 905 nm excitation. Highlighted emitters are single *kh* VVs.

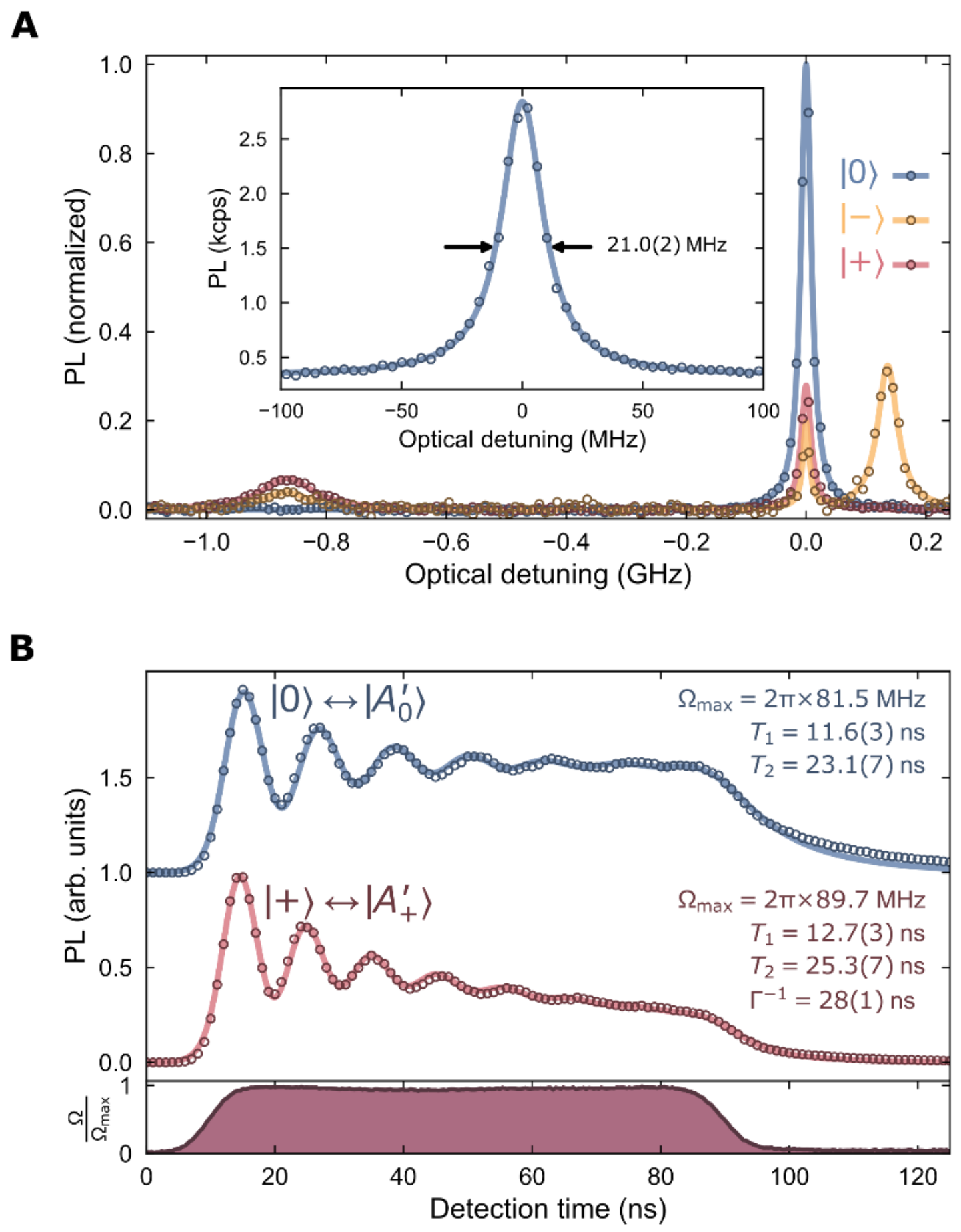

**Fig. 2. Optical properties of single *kh* VVs.** (**A**) Photoluminescence excitation spectrum of a single *kh* VV prepared into $|0\rangle$ (blue), $|-\rangle$ (yellow), and $|+\rangle$ (red) with 1 MV/m dc electric field applied and spectral diffusion compensated. Optical detuning measured with respect to 277.9337 THz (1078.647 nm). Inset: $|0\rangle \leftrightarrow |A_0'\rangle$ transition exhibiting a narrow, Lorentzian lineshape with spectral diffusion compensated. (**B**) Optical coherence of *kh* VVs. Optical Rabi oscillations between $|0\rangle \leftrightarrow |A_0'\rangle$ (blue circles) and $|+\rangle \leftrightarrow |A_+'\rangle$ (red circles) at 7.6 μW resonant excitation. Both transitions exhibit near-lifetime-limited optical coherence ($T_2 \approx 2T_1$). $|0\rangle \leftrightarrow |A_0'\rangle$ exhibits no detectable spin relaxation under illumination in this timescale, whereas excitation of $|+\rangle \leftrightarrow |A_+'\rangle$ rapidly depopulates $|+\rangle$. Bottom: The pulse envelope created by the acousto-optic modulator used to gate the resonant, narrow-line laser.

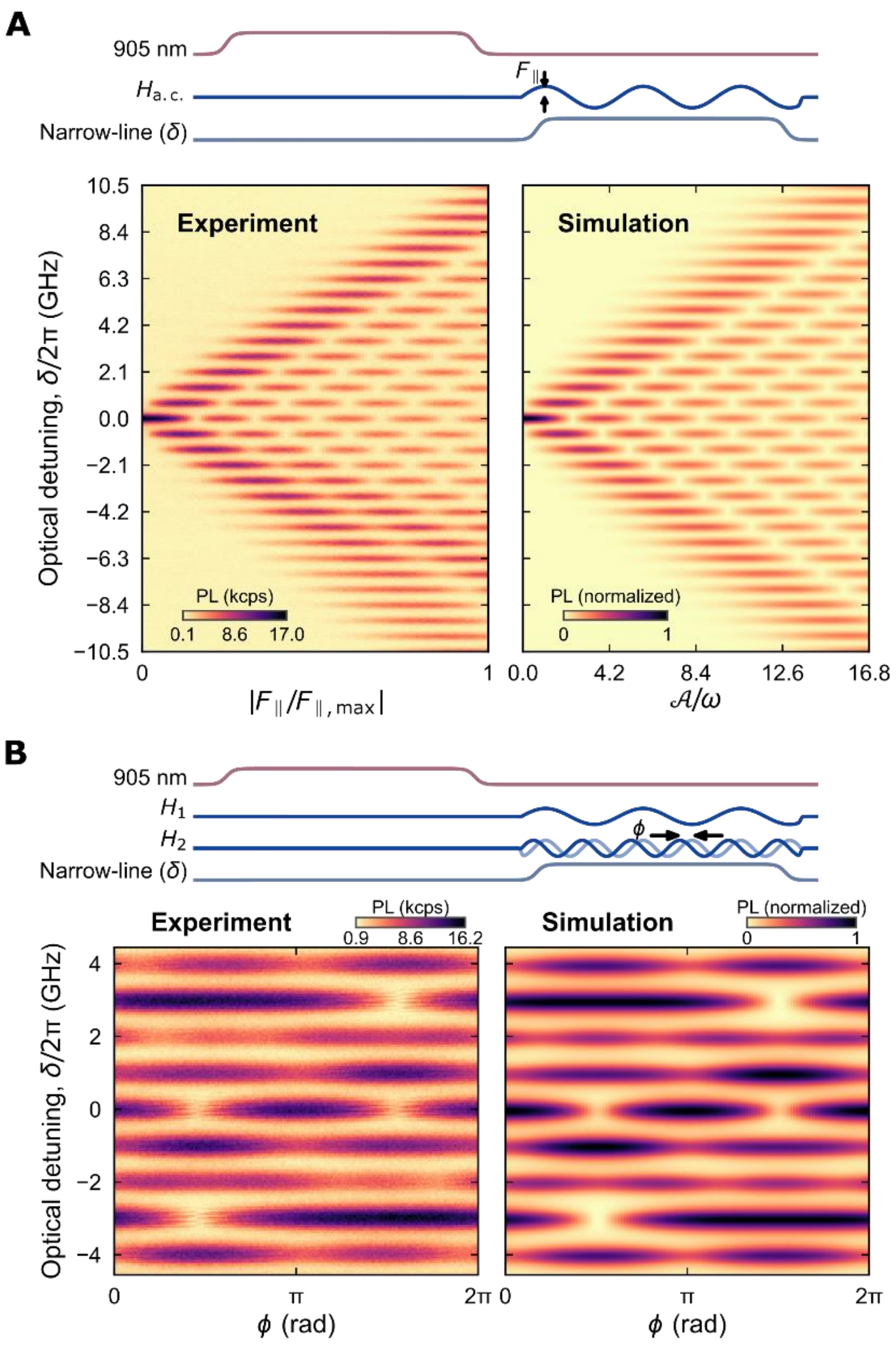

**Fig. 3. Landau-Zener-Stückelberg (LZS) interferometry of *kh* VV absorption spectrum** (**A**) Monochromatic LZS interferometry of *kh* VV absorption spectrum. Upper: Pulse sequence used to observe LZS interferometry. The interference pattern of $|0\rangle \leftrightarrow |A_0'\rangle$ is measured as a function of $\mathcal{A}$, the induced Stark shift amplitude. Bottom: Interference fringes of $|0\rangle \leftrightarrow |A_0'\rangle$ absorption arise in PLE spectroscopy as electric field magnitude $|\mathbf{F}|$ is increased ($|F_{\parallel,\max}| \approx 2$ MV/m), proportionally increasing $\mathcal{A}$. (**B**) Bichromatic LZS interferometry of *kh* VV absorption spectrum. Upper: Pulse sequence used to observe bichromatic LZS interferometry. The interference pattern of $|0\rangle \leftrightarrow |A_0'\rangle$ is measured as a function of the relative phase $\varphi$ of the two drives. Lower: PLE of a single *kh* VV under two electric field drives ($\omega_1 = 2\pi \times 1$ GHz, $\omega_2 = 2\pi \times 2$ GHz, $\mathcal{A}_1/\omega_1 = \mathcal{A}_2/\omega_2 \approx 2.4048$) as a function of $\varphi$. Multiphoton resonances arise at $1 \times n$ GHz and $2 \times n$ GHz optical detunings, resulting in fringes from constructive and destructive interference of the two drives.

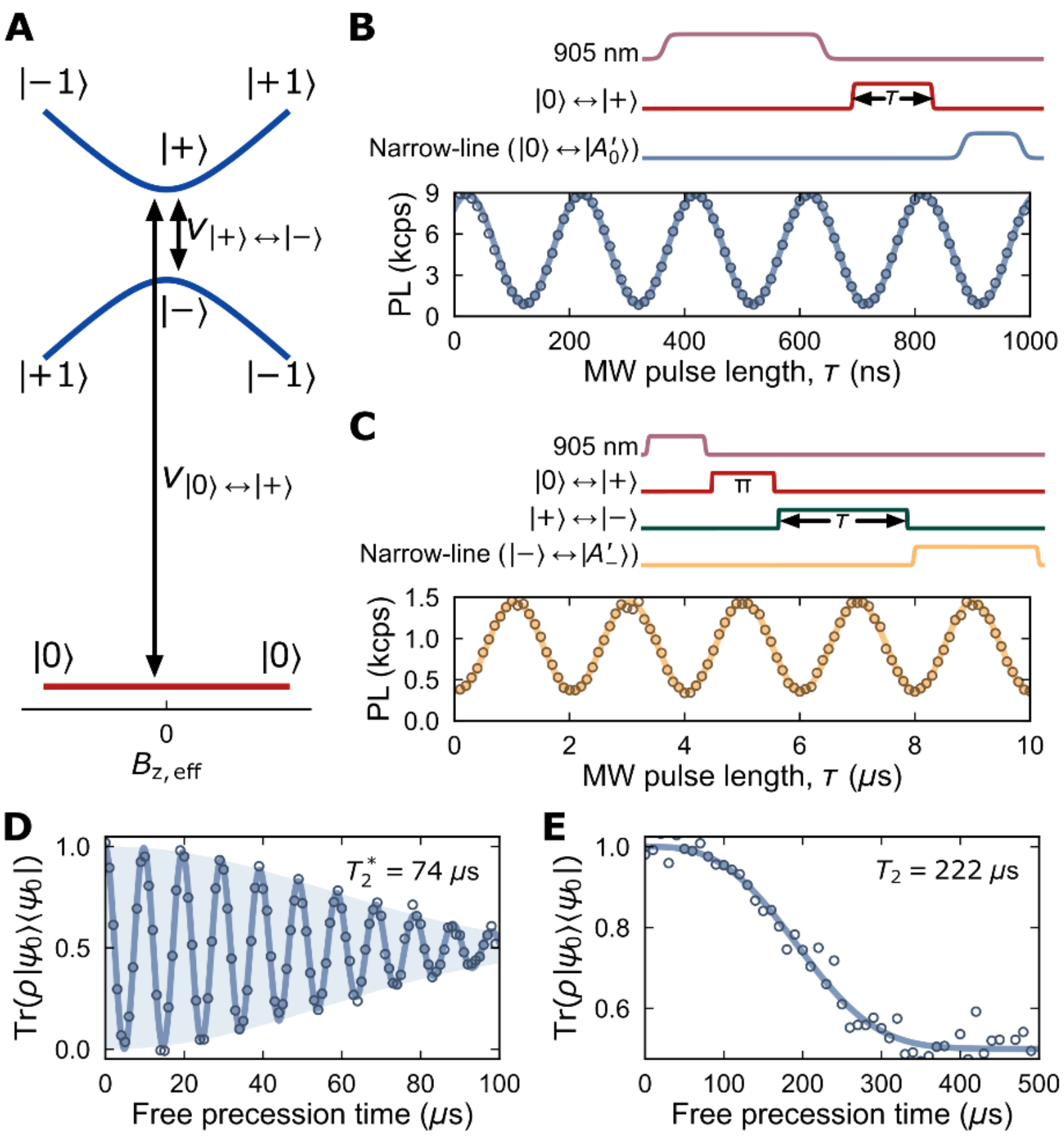

**Fig. 4. Near-ZEFOZ spin control and dynamics of single *kh* VVs.** (**A**) ZEFOZ transitions near zero effective magnetic field. Energy dispersion with respect to $B_z$ shows the vanishing first derivative of the spin transition energies, $\nu_{|0\rangle \leftrightarrow |\pm\rangle}$ and $\nu_{|+\rangle \leftrightarrow |-\rangle}$, at $B_{z,\text{eff}} = 0$. (**B**) Upper: Pulse sequence used to observe Rabi oscillations between |0⟩ and |+⟩. Lower: Rabi oscillations of the ground-state spin between |0⟩ and |+⟩. PL measured from $|0\rangle \leftrightarrow |A_0'\rangle$ excitation. (**C**) Upper: Pulse sequence used to observe Rabi oscillations between |+⟩ and |–⟩. Lower: Rabi oscillations of ground-state spin between |+⟩ and |–⟩. PL measured from $|-\rangle \leftrightarrow |A_-'\rangle$ excitation. The nearby $|0\rangle \leftrightarrow |A_0'\rangle$ transition increases background and reduces the contrast of Rabi oscillations. (**D**) Ramsey interferometry of a spin superposition prepared in $|\psi_0\rangle = \frac{1}{\sqrt{2}}(|0\rangle + |+\rangle)$. Dephasing mechanisms evolve the initial state $\rho(0) = |\psi_0\rangle\langle\psi_0|$ into $\rho(t)$. A microwave detuning of +100 kHz is added to increase visibility of the decay envelope. Readout is performed using $|0\rangle \leftrightarrow |A_0'\rangle$ PL. (**E**) Hahn-echo coherence of $|\psi_0\rangle = \frac{1}{\sqrt{2}}(|0\rangle + |+\rangle)$. A Gaussian decay envelope suggests the dominant source of spin decoherence is from the fluctuations of the $^{29}$Si and $^{13}$C nuclear spin bath.